\documentclass[aps,prd,twocolumn,showpacs,nofootinbib]{revtex4-2}
\usepackage{graphicx}
\usepackage{color}
\graphicspath{{figures/}{fig/}}
\usepackage{amsmath}
\usepackage{amssymb}
\usepackage{bm}
\usepackage{slashed}
\usepackage{epsfig}
\usepackage{amsfonts}
\usepackage{epstopdf}
\usepackage{hyperref}
\usepackage{bbm}
\usepackage{textcomp}
\usepackage{color}
\usepackage[ruled,linesnumbered]{algorithm2e}

\newcommand{\sect}[1]{\section{#1}}
\newcommand{\mi}{\mathrm{i}}
\newcommand{\md}{\mathrm{d}}

\begin{document}

\title{ Exploring the linear space of Feynman integrals via generating functions}

\author{Xin Guan}
\email{guanxin0507@pku.edu.cn}
\affiliation{School of Physics, Peking University, Beijing 100871, China}

\author{Xiang Li}
\email{lix-PHY@pku.edu.cn}
\affiliation{School of Physics, Peking University, Beijing 100871, China}

\author{Yan-Qing Ma}
\email{yqma@pku.edu.cn}
\affiliation{School of Physics, Peking University, Beijing 100871, China}
\affiliation{Center for High Energy Physics, Peking University, Beijing 100871, China}

\begin{abstract}
Deriving a comprehensive set of reduction rules for Feynman integrals has been a longstanding challenge. In this paper, we present a proposed solution to this problem utilizing generating functions of Feynman integrals. By establishing and solving differential equations of these generating functions, we are able to derive a system of reduction rules that effectively reduce any associated Feynman integrals to their bases. We illustrate this method through various examples and observe its potential value in numerous scenarios.
\end{abstract}

\maketitle

\sect{Introduction}
Scattering amplitudes play a crucial role in quantum field theory as they connect theoretical predictions with experimental observations. With the successful operation of the Large Hadron Collider \cite{Azzi:2019yne,Cepeda:2019klc} and the proposal of next-generation colliders\cite{ILC:2013jhg,Behnke:2013xla,Bambade:2019fyw,CEPCStudyGroup:2018ghi,CEPCStudyGroup:2018rmc,TLEPDesignStudyWorkingGroup:2013myl,FCC:2018byv,FCC:2018evy}, perturbative calculations of scattering processes need to be pushed to higher orders, such as next-to-next-to-leading order, to match the precision of experimental measurements. This requirement necessitates the computation of scattering amplitudes at multiloop levels. Utilizing Lorentz symmetry, these amplitudes can be expressed as linear combinations of scalar Feynman integrals (FIs). The calculation of scalar FIs poses a significant challenge for state-of-the-art problems.

A family of scalar FIs can be represented as
\begin{align}\label{eq:ampara0}
I(\vec{\nu})=\int\prod_{i=1}^{L}\frac{\md^{D}\ell_i}{\mi\pi^{D/2}}
\frac{\mathcal{D}_{K+1}^{-\nu_{K+1}}\cdots \mathcal{D}_N^{-\nu_N}}{\mathcal{D}_1^{\nu_{1}}\cdots \mathcal{D}_K^{\nu_{K}}},
\end{align}
where $D = 4-2\epsilon$ represents the spacetime dimension, $L$ is the number of loops, $\ell_i$ are the loop momenta, $\mathcal{D}_1,\ldots,\mathcal{D}_K$ are the inverse propagators, $\mathcal{D}_{K+1},\ldots,\mathcal{D}_N$ are the irreducible scalar products (ISPs) introduced for completeness, $\nu_1,\ldots,\nu_K$ can be any integers, and $\nu_{K+1},\ldots,\nu_N$ can only be non-positive integers. The \emph{rank} of an integral is defined as the opposite value of the sum of all negative powers, and the \emph{dots} of an integral represent the sum of all positive powers subtracted by the number of positive indices. For instance, the integral $I(1,2,2,3,-1,-2)$ has rank 3 and dots 4.

It has been proven that a family of FIs forms a finite-dimensional linear space \cite{Smirnov:2010hn} with bases known as  master integrals (MIs). Therefore, the prevailing method for calculating scalar FIs involves two distinct tasks. The first task is FIs reduction, which aims to express FIs as linear combinations of MIs \cite{Chetyrkin:1981qh, Gehrmann:1999as, Laporta:2000dsw, Baikov:2005nv, Baikov:2007zza, Gluza:2010ws,Schabinger:2011dz, Larsen:2015ped, Wu:2023upw, vonManteuffel:2014ixa, Peraro:2016wsq,  Peraro:2019svx, Klappert:2019emp, Klappert:2020aqs,  Belitsky:2023qho, Liu:2018dmc, Guan:2019bcx, www:Blade, Smirnov:2020quc,Usovitsch:2020jrk, Anastasiou:2004vj, Smirnov:2008iw,Smirnov:2013dia,Smirnov:2014hma,Smirnov:2019qkx, Maierhofer:2017gsa, Maierhofer:2018gpa,Klappert:2020nbg, Lee:2012cn, Lee:2013mka, Studerus:2009ye,vonManteuffel:2012np,   Mastrolia:2018uzb, Frellesvig:2019kgj,  Frellesvig:2020qot,Weinzierl:2020xyy, Wang:2019mnn, Boehm:2020ijp, Basat:2021xnn, Heller:2021qkz, Bendle:2021ueg}; While the second is to compute these MIs \cite{Hepp:1966eg, Roth:1996pd, Binoth:2000ps, Heinrich:2008si, Smirnov:2008py,Smirnov:2009pb,Smirnov:2013eza,Smirnov:2015mct,Smirnov:2021rhf, Carter:2010hi, Borowka:2012yc, Borowka:2015mxa, Borowka:2017idc, Borowka:2018goh, Bergere:1973fq, Boos:1990rg, Smirnov:1999gc, Tausk:1999vh, Czakon:2005rk, Smirnov:2009up, Gluza:2007rt, www:mbtools, Belitsky:2022gba, Beneke:1997zp, Jantzen:2012mw, Lee:2009dh, Kotikov:1990kg, Kotikov:1991pm, Remiddi:1997ny, Gehrmann:1999as, Argeri:2007up, Muller-Stach:2012tgj, Henn:2013pwa, Henn:2014qga, Moriello:2019yhu, Hidding:2020ytt, Armadillo:2022ugh, Catani:2008xa, Bierenbaum:2010cy, Bierenbaum:2012th, Tomboulis:2017rvd, Runkel:2019yrs, Capatti:2019ypt, Aguilera-Verdugo:2020set, brown:2009aaa, Panzer:2015ida, Panzer:2014caa, Hidding:2022ycg, Liu:2017jxz,Liu:2020kpc, Liu:2021wks,Liu:2022chg,Liu:2022tji,Liu:2022mfb,Song:2021vru,Zeng:2023jek}. Notably, based on the auxiliary mass flow method \cite{Liu:2017jxz,Liu:2020kpc, Liu:2021wks,Liu:2022chg,Liu:2022tji,Liu:2022mfb}, any given FI can be automatically calculated to high precision as far as the reduction has been achieved. 
However, FIs reduction is a critical yet formidable task in complicated multiloop processes.

Integration-by-parts (IBP) identities \cite{Chetyrkin:1981qh} combining with the Laporta algorithm \cite{Laporta:2000dsw} is a widely-used approach  to realize reduction.  Even though powered by the finite field method \cite{vonManteuffel:2014ixa, Peraro:2016wsq,  Peraro:2019svx, Klappert:2019emp, Klappert:2020aqs,  Belitsky:2023qho} to avoid intermediate expression swell, and syzygy equations\cite{Gluza:2010ws,Schabinger:2011dz,Cabarcas:2011ddd,Larsen:2015ped,Ita:2015tya,Zhang:2016kfo,Bohm:2018bdy,vonManteuffel:2020vjv,Badger:2021imn,Abreu:2021asb} and block-triangular form \cite{Liu:2018dmc,Guan:2019bcx,www:Blade} to reduce the size of IBP system, reduction of  FIs with high power of denominators or ISPs is still extremely time- and resource-consuming. Various other methods have been proposed to bypass IBP, but each approach has its own difficulties. For instance, in the method based on intersection theory \cite{Mastrolia:2018uzb, Frellesvig:2019kgj,  Frellesvig:2020qot,Weinzierl:2020xyy}, the calculation of intersection numbers for multi-variable problems remains challenging; In the methods based on large spacetime expansion \cite{Baikov:2005nv,Baikov:2007zza} or large auxiliary mass expansion \cite{Liu:2018dmc}, it is hard to obtain very higher order expansion terms.

For a long time, the ultimate goal of FIs reduction has been to find a complete set of reduction rules, known as \emph{recurrence relations}, with general powers $\vec{\nu}$. These recurrence relations should efficiently reduce all integrals in a family to the MIs. 
For simple problems, such recurrence relations can be constructed by analyzing IBP identities manually, see e.g Refs. \cite{Chetyrkin:1981qh,Tkachov:1983bak,Gray:1990yh,Broadhurst:1991fz,Davydychev:1992mt,Caffo:1998du,Melnikov:2000qh} for early works. There are also very powerful reduction programs for specific problems, which can tackle massive tadpoles, like \texttt{MATAD} \cite{Steinhauser:2000ry}, massless self-energy topoly at three and four loop, respectively, called \texttt{MINCER} \cite{Gorishnii:1989gt,Larin:1991fz} and \texttt{FORCER} \cite{Ruijl:2017cxj}. A tailored heuristical approach for general problems is also available in \texttt{LiteRed} \cite{Lee:2013mka}. However, these construction procedures are obscure and success is not guaranteed. 

From an algebraic geometry perspective, the establishment of recurrence relations becomes possible once the Gr\"obner bases are known. Gr\"obner bases serve as a fundamental tool in the analysis of polynomial ideals, and significant efforts have been dedicated to this area of research
\cite{Tarasov:2004ks,Gerdt:2005qf,Smirnov:2005ky,Smirnov:2006tz,Smirnov:2006wh,Lee:2008tj,Barakat:2022qlc}. However, computing Gr\"obner bases for the noncommutative algebra generated by the IBP relations remains an exceedingly challenging task.

In this paper, we propose a general approach to deriving recurrence relations for arbitrary Feynman integral families by setting up and solving differential equations (DEs) for generating functions (GFs) associated with these integrals. By doing so, we are able to effectively reduce the infinite number of Feynman integrals within a family to a finite set of MIs. This reduction is achieved through solving a linear system with finite size, thereby providing a solution to this long-standing and challenging problem with reasonable computational complexity.

We employ the recently released \texttt{Blade} package \cite{www:Blade} to construct DEs of GFs. To demonstrate the performance of the GF method, we apply it to some  examples, 
spanning from simple and straightforward problems to cutting-edge and complex scenarios.

\sect{Generating functions}\label{sec:GF}

Our approach is motivated by the auxiliary mass flow method \cite{Liu:2018dmc}, where by introducing an auxiliary mass term $\eta$ to each denominator and expanding around $\eta=0$ using DEs w.r.t. $\eta$, one can obtain values of a series of FIs with arbitrary powers of denominators.  It means that the FI with auxiliary masses serves as a GF of these original FIs. Similar ideas have also been presented in Ref. \cite{Tarasov:2004ks}, where Gr\"obner bases are computed by solving DEs when all propagators, including ISPs, have different masses. In Ref. \cite{Ablinger:2012sm}, the summation of the non-standard term $(\Delta \cdot p)^n$ into a linear propagator $1/(1-x \Delta \cdot p)$ constructs a kind of GF and has been used in\cite{Blumlein:2021enk, Blumlein:2021ryt, Gehrmann:2023ksf} for reduction of high rank integrals. In Refs. \cite{Feng:2022hyg}, GFs are employed to reduce arbitrary one-loop tensor integrals by introducing an auxiliary vector \cite{Feng:2021enk,Hu:2021nia,Feng:2022iuc,Feng:2022rfz,Feng:2022uqp}. In this paper, we propose to use GF method to reduce all FIs in any given family.

One choice of GFs for the integral family in Eq.~\eqref{eq:ampara0} can be defined as follows:
\begin{align} \label{eq:expGF}
\begin{split}
G_{\vec{\nu}}(\vec{x},\vec{\eta})&=\int\prod_{i=1}^{L}\frac{\md^{D}\ell_i}{\mi\pi^{D/2}}  \frac{e^{Z^{\vec{x}}} \prod_{i=K+1}^N{\cal D}_{i}^{-\nu_i} }{ \prod_{i=1}^K({\cal D}_{i}-\eta_{i})^{\nu_i} }, 
\end{split}
\end{align}
with top-sector corner integral given by
\begin{align} \label{eq:expGFtop}
\begin{split}
G_{{\text{top}}}(\vec{x},\vec{\eta})&=\int\prod_{i=1}^{L}\frac{\md^{D}\ell_i}{\mi\pi^{D/2}}  \frac{e^{Z^{\vec{x}}} }{ \prod_{i=1}^K({\cal D}_{i}-\eta_{i}) }, 
\end{split}
\end{align}
where ${\nu}_i$ are given integers, $\eta_i$ are auxiliary masses and $Z^{\vec{x}}={\sum_{i=K+1}^N x_i{\cal D}_i}$
is a linear combination of ISPs. FIs in Eq.~\eqref{eq:ampara0} can be generated from $G_{{\text{top}}}(\vec{x},\vec{\eta})$ by expanding $\eta_i$ around $0$ or $\infty$ and $x_i$ around $0$. For example, FIs in the top-sector can be generated by taking the following limit: \footnote{As in general $\eta_i=0$ and $ x_i=0$ are singular points of $G$,  the limit $\vec{\eta}\rightarrow0$ and $\vec{x}\rightarrow0$ means selection of the Taylor branch of the GF and FIs are coefficients of Taylor expansion of GFs.
Because each branch of GFs satisfies the same DEs w.r.t. $\vec{\eta}$ and $\vec{x}$, we will not emphasize the Taylor branches in the rest of the paper.
}
\begin{align}
I(\vec{\nu})=\lim_{\vec{x}\rightarrow0}\lim_{\vec{\eta}\rightarrow0} \partial^{\vec{\nu}}_{x} \partial^{\vec{\nu}}_{\eta}  G_{{\text{top}}}(\vec{x},\vec{\eta})
\end{align}
where
\begin{align}
\partial^{\vec{\nu}}_{x}&\equiv\prod_{i=K+1}^{N}\frac{\partial^{\nu_i}}{{\partial x_i}^{\nu_i}},\\
\partial^{\vec{\nu}}_{\eta}&\equiv \prod_{i=1}^{K}\frac{1}{{(\nu_i-1)}! }\frac{\partial^{\nu_i-1}}{{\partial \eta_i}^{\nu_i-1}}.
\end{align}

Similar to FIs, using methods such as IBP method, we find that the GFs defined in Eq.~\eqref{eq:expGF} form a finite-dimensional linear space. We can choose a set of MIs of GFs that cover $G_{{\text{top}}}(\vec{x},\vec{\eta})$, denoted as $\vec{\cal G}(\vec{x},\vec{\eta})$,  and derive DEs for them:
\begin{align}
\frac{\partial }{\partial x_i}\vec{\cal G}=& A_i \vec{\cal G},\\
\frac{\partial }{\partial \eta_i}\vec{\cal G}=& B_i \vec{\cal G}.
\end{align}
Since the FIs are expansion coefficients of GFs, by expanding  $\vec{\cal G}(\vec{x},\vec{\eta})$ using DEs, we can obtain a system of linear relations between FIs. These relations are the desired recurrence relations that efficiently reduce all target FIs to the minimal set of MIs. Therefore, once we obtain the DEs for the MIs of GFs, the reduction of FIs in the corresponding family is solved.

In physical Feynman amplitudes, one often encounters FIs with high rank but very small dots. In such cases, we can define simpler GFs as follows:
\begin{align} \label{eq:expGF0}
\begin{split}
G_{\vec{\nu}}(\vec{x})&=\int\prod_{i=1}^{L}\frac{\md^{D}\ell_i}{\mi\pi^{D/2}}  \frac{e^{Z^{\vec{x}}}}{\prod_{i=1}^N {\cal D}_{i}^{\nu_i}} ,
\end{split}
\end{align}
and corresponding simpler MIs $\vec{\cal G}(\vec{x})$, which have fewer extra variables. By setting up and solving partial DEs of $\vec{\cal G}(\vec{x})$ w.r.t. $x_i$, we can achieve the reduction of original FIs with arbitrary rank. 

However, in practice, we find that there is an even better approach, denoted as fixed direction scheme, by setting
\begin{align}
    \vec{x}= x \vec{c},
\end{align}
where $x$ is a variable and $\vec{c}$ are some fixed numbers. With each chosen set of values for $\vec{c}$, we define GFs
\begin{align} \label{eq:expGFx}
\begin{split}
G_{\vec{\nu}}^{\vec{c}}({x})&=\int\prod_{i=1}^{L}\frac{\md^{D}\ell_i}{\mi\pi^{D/2}}  \frac{e^{x Z^{\vec{c}}}}{\prod_{i=1}^N {\cal D}_{i}^{\nu_i}} ,
\end{split}
\end{align}
and set up DEs for the corresponding MIs $\vec{\cal G}^{\vec{c}}(x)$ w.r.t. $x$,
\begin{align}
\frac{\partial }{\partial x}\vec{\cal G}^{\vec{c}}=& A^{\vec{c}} \vec{\cal G}^{\vec{c}}.
\end{align}
By expanding $\vec{\cal G}^{\vec{c}}(x)$ around $x=0$ using the above DEs, we can obtain a system of relations between FIs. By simulating a sufficient number of $\vec{c}$ values, we can generate enough relations to reduce the original FIs of a certain rank. One advantage of the fixed direction scheme is that there are only two singularities in $x$ space, namely $x=0$ and $\infty$, which makes the DEs w.r.t. $x$ relatively simple and easier to achieve.

It is important to note that there are many other possible choices of GFs. For example, one possibility is to replace $e^{Z^{\vec{x}}}$ by $1/(1-Z^{\vec{x}})$. However, we find that this choice is generally less efficient because the singularities in the $\vec{x}$-plane become more complicated.

\begin{figure}[h]
    \includegraphics[width=5cm]{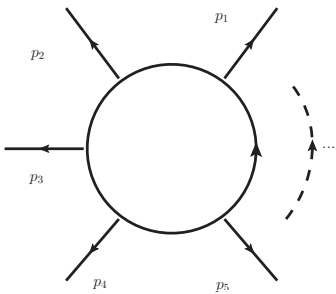}
    \caption{One-loop N-point diagram.}
    \label{fig:1loop}
\end{figure}

\sect{One-loop N-point}\label{sec:oneloop}
For a general one-loop N-point family shown in Fig.\ref{fig:1loop}, we aim to achieve a comprehensive set of reduction rules with general powers of propagators through the method of GFs. The obtained recurrence relations are not new, which has been derived previously, e.g. in Ref. \cite{Duplancic:2003tv}.

We consider a set of external momenta ${k_1,\dots,k_N}$ which satisfy momentum conservation $\sum_{i=1}^N k_i =0$. Inverse propagators can be defined as
\begin{align}
\mathcal{D}_i={(\ell+r_i)}^2-m_i^2,\quad i=1,\dots, N\ ,
\end{align}
where $\ell$ is loop momentum, the momenta $r_i$ are defined by $r_i = k_i + r_{i-1}$ for $i = 1, \dots, N$, and $r_0 = r_N=0$ by definition. For the sake of clarity, we omit the infinitesimal imaginary part $\mi \epsilon$ in inverse propagators, since the reductions are independent of it.
As there is no ISP at the one-loop level, we introduce GFs as
\begin{align}
G_{\vec{\nu}}(\vec{\eta})=\int\frac{\md^{D}\ell}{\mi\pi^{D/2}}\frac{1}{ \prod_{i=1}^N ({\cal D}_{i}-\eta_{i})^{\nu_i} },
\end{align}
and MIs of GFs can be chosen as that with $\nu_i$ being $0$ or $1$.
The IBP equations 
\begin{align}
    \int \frac{\md^{D}\ell}{\mi\pi^{D/2}} \frac{\partial}{\partial \ell^{\mu}}\frac{\ell^{\mu} + r_i^{\mu}}{\prod_{j=1}^N({\cal D}_{j}-\eta_{j})} = 0,\quad i=1,\dots,N\ ,
\end{align}
lead to DEs of GFs:
\begin{align}\label{eq:1looprelation}
    \begin{split}
        &\sum_{j=1}^N (R_{j i} - \eta_j-\eta_i) \frac{\partial G_{\text{top}}}{\partial \eta_j} \\
        &= \sum_{j\neq i}^N \frac{\partial G_{\text{top}-\vec{e}_i}}{\partial \eta_j}+(1-D+N) G_{\text{top}},
    \end{split}
\end{align}
where $R_{ji} = (r_{j} - r_{i})^2 - m_i^2 - m_j^2$ is the Gram matrix of the one-loop N-point family, 
$\vec{e}_i$ with $i=1,\dots,N$ are unit vectors and ``$\text{top}-\vec{e}_i$" means ``$(1,\cdots,1)-\vec{e}_i$".

By expanding Eq.\eqref{eq:1looprelation} at small value of $\vec{\eta}$, the coefficients of $\prod_{k=1}^N\eta_k^{\nu_k-1}$ result in
\begin{align}\label{eq:1looprecursion}
    \begin{split}
        &\sum_{j=1}^N R_{ji} \nu_j I(\{\nu_k + \delta_{kj}\}) =\\
        &\sum_{j=1}^N \nu_j I(\{\nu_k + \delta_{kj} - \delta_{ki}\}) - (D-\sum_{j=1}^N \nu_j)I(\{\nu_k\}),
    \end{split}
\end{align}
which can reduce all FIs with nonzero dots to MIs. Therefore, we find that the DEs of GFs provide all desired recurrence relations at one-loop level. Our result is consistent with that presented in Eq.(1) of Ref. \cite{Duplancic:2003tv}. 

Furthermore, by expanding Eq.\eqref{eq:1looprelation} at small value of $1/\eta_i$ for some index $i$ and small value of $\eta_i$ for the others, we can gain the same recurrence relations in Eq. \eqref{eq:1looprecursion}, which can reduce FIs with negative powers. FIs with negative powers can also be viewed as tensor integrals, which can be related to scalar integrals via Lorentz decomposition \cite{Passarino:1978jh}. Alternatively, the tensor integrals can also be reduced by GFs with auxiliary vector \cite{Feng:2022hyg}.

\begin{figure}[h]
    \includegraphics[width=4.5cm]{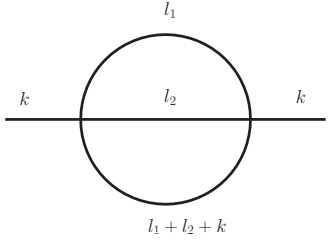}
    \caption{Two-loop sunrise diagram.}
    \label{fig:sunrise}
\end{figure}

\sect{Two-loop sunrise}\label{sec:sunrise}
Now let us give a simple two-loop example, the massless sunrise family shown in Fig.\ref{fig:sunrise}.
The family has two loop momenta $\ell_1$ and $\ell_2$ and one external momenta $k$, satisfying $k^2=s$. We choose a complete set of Lorentz scalars as
\begin{align}
\begin{split}
    &\mathcal{D}_1={\ell_1}^2, \mathcal{D}_2={\ell_2}^2, \mathcal{D}_3={(\ell_1+\ell_2+k)}^2,\\
    &\mathcal{D}_4= \ell_1 \cdot k, \mathcal{D}_5 = \ell_2 \cdot k,
\end{split}
\end{align}  
where the last two are ISPs.
We choose the ISPs as products of loop momenta and external momenta, instead of quadratic forms like $(\ell_1+k)^2$, since the former have simpler DEs in practice.

Although we can construct DEs w.r.t $x_i$ and $\eta_i$ to obtain recurrence relations to reduce all FIs in the family, this is not necessary. Instead, reduction problems can be usually divided into two cases: 1)  with large rank but small dots; 2) with large dots but small rank. We thus will only deal with these two cases.

Considering arbitrary powers of ISPs, we choose GFs defined in Eq.~\eqref{eq:expGF0}:
\begin{align}
G_{\vec{\nu}}(\vec{x})=\int\frac{\md^{D}\ell_1 \md^{D}\ell_2}{(\mi\pi^{D/2})^2}
\frac{\mathcal{D}_{4}^{-\nu_{4}}\mathcal{D}_{5}^{-\nu_{5}} e^{Z^{\vec{x}}}}{\mathcal{D}_1^{\nu_{1}} \mathcal{D}_2^{\nu_{2}} \mathcal{D}_3^{\nu_{3}}}, 
\end{align}
where ${\nu_4, \nu_5}$ are non-positive integers only.
MIs can be chosen as
\begin{align}
    \vec{\cal{G}} = \left(G_{(1,1,1,0,0)}, G_{(1,1,1,0, -1)}, G_{(1,1,1,-1, 0)}\right),
\end{align}
which include only top-sector integrals because all sub-sector integrals in this family are scaleless and vanishing in dimensional regularization.
Using \texttt{Blade}, we get the DEs w.r.t. $x_i$: 
\begin{align} \label{eq:sunriseDEs}
    \frac{\partial}{\partial x_i} \vec{\cal{G}} = A_i \; \vec{\cal{G}},
\end{align}
where
\begin{align}
    A_4 = \begin{pmatrix}
        0 & 1 & 0 \\
        \frac{(-1+\epsilon)s}{x_4} & \frac{(-1+\epsilon)(3 x_4 - 2 x_5)}{x_4 (x_4 - x_5)} - \frac{s}{2} & -\frac{(-1+\epsilon) x_5}{x_4 (x_4 - x_5)}\\
        -\frac{s^2}{4} & -\frac{-1+\epsilon}{x_4 - x_5} -\frac{s}{2} & \frac{-1+\epsilon}{x_4 - x_5} -\frac{s}{2}
    \end{pmatrix},
\end{align}
\begin{align}
    A_5 = \begin{pmatrix}
        0 & 0 & 1 \\
        -\frac{s^2}{4} & -\frac{-1+\epsilon}{x_4 - x_5} -\frac{s}{2} & \frac{-1+\epsilon}{x_4 - x_5} -\frac{s}{2}\\
        \frac{(-1+\epsilon)s}{x_5} & \frac{(-1+\epsilon) x_4}{x_5 (x_4 - x_5)} & \frac{(-1+\epsilon)(2 x_4 - 3 x_5)}{x_5 (x_4 - x_5)} - \frac{s}{2}
    \end{pmatrix},
\end{align}

By expanding the DEs to $x_4^{\nu_4}x_5^{\nu_5}$ with arbitrary $\nu_4$ and $\nu_5$, we obtain three different linear relations between FIs.  Due to the permutation symmetry of the sunrise FIs, $I(\nu_4,\nu_5)\equiv I(1,1,1,\nu_4,\nu_5)=I(1,1,1,\nu_5,\nu_4)$, we only need to consider the cases where $\nu_4\leq\nu_5\leq0$ and thus we have the following relation:
\begin{align} \label{eq:sunriserec}
    \begin{split}
        &c_0 I(\nu_4,\nu_5)\\
    =&c_1 I(1 + \nu_4, \nu_5) + c_2 I(1 + \nu_4, 1 + \nu_5) + c_3 I(2 + \nu_4, \nu_5),
    \end{split}
\end{align}
where
\begin{align*}
    \begin{split}
    c_0&=4 (-1 + 2 \epsilon + \nu_4)[(\nu_5-1)(\nu_4 + \nu_5-2)\\
    &+ \epsilon (\nu_4^2 - \nu_5^2 + 7 \nu_5 - 7)+3 \epsilon^2 (2 + \nu_4 - \nu_5)],\\ 
    c_1&= -2s[(\nu_4^3 + \nu_4^2 (2 \nu_5-3)+\nu_4 (1 - \nu_5) + (\nu_5-1)^2)\\
    &+ \epsilon (\nu_4^3 + \nu_4^2 (8 - \nu_5) +2 \nu_4 (3 \nu_5-5)-(\nu_5-1) (2 \nu_5-5))\\
    &+\epsilon^2 (2 \nu_4^2 + \nu_4 (19 - 3 \nu_5) + (5-\nu_5)(1-\nu_5))\\
    &+\epsilon^3 2(1 - \nu_4 + \nu_5)],\\
    c_2&=-s^2(-1 + \epsilon) (-1 + 2 \epsilon + \nu_4) \nu_5 (-1 + 2 \epsilon + \nu_5),\\
    c_3&=-s^2(1 + \nu_4) (2 \epsilon + \nu_4) ( \nu_4+\nu_5 -2 - \epsilon(\nu_5-5) - 2 \epsilon^2),\\
    \end{split}
\end{align*}
This recurrence relation can reduce $I(\nu_4,\nu_5)$ with $\nu_4\leq\nu_5\leq 0$ to $I(0,0)$.
Note that, as Eq.~\eqref{eq:sunriserec} is an analytic function of $\nu_4$ and $\nu_5$, it also holds even if $\nu_4$ and $\nu_5$ are positive.

To obtain the reduction of arbitrary powers of propagators with no ISP, we introduce GFs:
\begin{align}
\begin{split}
&G_{\vec{\nu}}(\vec{\eta})=\\
&\int\frac{\md^{D}\ell_1 \md^{D}\ell_2}{(\mi\pi^{D/2})^2}
\frac{1}{(\mathcal{D}_1-\eta_1)^{\nu_1} (\mathcal{D}_2-\eta_2)^{\nu_2}(\mathcal{D}_3-\eta_3)^{\nu_3}}.
\end{split}
\end{align}
MIs can be effectively chosen as
\begin{align*}
    \vec{\cal{G}} = \left(G_{(1,1,1)}, G_{(2,1,1)},G_{(1,2,1)}, G_{(1,1,2)}\right),
\end{align*}
because the expansion coefficients of $G_{\vec{\nu}}(\vec{\eta})$ are scaleless if any $\nu_i \leq 0$.
The DEs of $\vec{\cal{G}}$ w.r.t. $\eta_i$ and the corresponding recurrence relations can be similarly obtained, which are available in the ancillary file.

\sect{Two-loop double-box}\label{sec:double-box}

\begin{figure}[h]
    \includegraphics[width= 4cm]{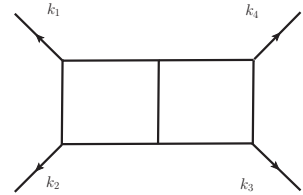}
    \caption{Two-loop double-box diagram.}
    \label{fig:doublebox}
\end{figure}

In this section, we take the double-box diagram shown in Fig.\ref{fig:doublebox} as an example to demonstrate that: 1) the fixed direction scheme is usually preferred;  and 2) the construction of DEs of GFs is usually much easier than the construction of recurrence relations of original FIs directly. 

In this problem, there are four external momenta $k_1$, $k_2$, $k_3$, $k_4$ satisfying on-shell conditions $k_i^2=0$ and momentum conservation $\sum_{i=1}^4 k_i =0$, which leaves two independent scales $s=2 k_1 \cdot k_2$ and $t=2 k_2 \cdot k_3$. We choose a complete set of Lorentz scalars as
\begin{align}
\begin{split}
    &\mathcal{D}_1={\ell_1}^2,\ \mathcal{D}_2=(\ell_1+k_1)^2,\ \mathcal{D}_3={(\ell_1+k_1+k_2)}^2,\\
    &\mathcal{D}_4= {\ell_2}^2,\ \mathcal{D}_5= {(\ell_2-\ell_1)}^2,\ \mathcal{D}_6={(\ell_2+k_1+k_2)}^2,\\
    &\mathcal{D}_7={(\ell_2-k_4)}^2,\ \mathcal{D}_8 = \ell_1 \cdot k_3,\ \mathcal{D}_9 = \ell_2 \cdot k_1,
\end{split}
\end{align}
where the last two are ISPs. DEs of GFs in this family are too long to present, and they are available in the ancillary file.

\subsection{Fixed direction scheme}

Based on GFs defined in Eq.~\eqref{eq:expGF0}, there are 108 MIs with 5 in the top-sector.
We construct DEs of these MIs w.r.t. variables $x_i$ ($i=8,9$),
\begin{align} \label{eq:dbx89}
    \frac{\partial }{\partial x_i}\vec{\cal G}=& A_i \vec{\cal G} \ ,
\end{align}
and find that there are many singularities in the coefficient matrices.
For example, the matrix elements $(A_8)_{6,6}$, associated with $\partial_{x_8} G_{(0, 0, 1, 1, 1, 0, 0, -1, -1)}$ and $G_{(0, 0, 1, 1, 1, 0, 0, -1, -1)}$, possesses a denominator
\begin{align}
    \text{den}((A_8)_{6,6}) =&2 x_8 \times(s x_8^2 + 2 s x_8 x_9 + 2 t x_8 x_9 + s x_9^2)\nonumber\\
    \times&(s x_8^2 + 2 s x_8 x_9 + 4 t x_8 x_9 + s x_9^2) \ .
\end{align}
The complicated denominators lead to complicated numerators, and thus large number of numeric sample points are required to reconstruct the matrix $A_i$. Therefore, it is very time-consuming.

Alternatively, we can choose the fixed direction scheme defined in Eq.~\eqref{eq:expGFx}. In this scheme there are the same number of MIs as the previous scheme, but DEs w.r.t  $x$,
\begin{align}
    \frac{\partial }{\partial x}\vec{\cal G}^{\vec{c}}=& A^{\vec{c}} \vec{\cal G}^{\vec{c}},
\end{align}
are much simpler. For example, by choosing $c_8 = 1$ and $c_9 =2$, the denominator of the matrix element $(A^{(1,2)}_{x})_{6,6}$ becomes
\begin{align}
    \text{den}((A^{(1,2)}_x)_{6,6}) = 2x(9 s + 4 t),
\end{align}
which results in also simpler numerators.
In fact, for any given value of $\vec{c}$, one can always choose  MIs of GFs \footnote{One can use the algorithms presented in Refs.\cite{Smirnov:2020quc, Usovitsch:2020jrk}. The original purpose of these algorithms is to choose proper MIs so that no coupled singularities between spacetime dimension $D$ and kinematic variables presenting in reduction coefficients.} so that $A^{\vec{c}}$ only has singularities at $x=0$ and $x=\infty$, and thus the construction of $A^{\vec{c}}$ is much easier than that of $A_i$ 
 in Eq.~\eqref{eq:dbx89}. As shown in Tab. \ref{tab:scheme}, construction of $A_i$, which are functions of $s,~t,~\epsilon,~x_8$ and $x_9$, needs 130 seconds CPU time; while the construction of $A^{\vec{c}}$ with given value of $\vec{c}$, which are functions of $s,~t,~\epsilon$ and $x$, needs 13 seconds.

\begin{table}[htbp]
    \begin{tabular}{|c|c|c|c|c|c|}
    \hline
         Scheme& \#MIs & $t_{\text{ibp}}$(s) & Points & Primes & $t_{\text{total}}$(s)\\
         \hline
         all&108 & 0.02 & 3307 & 2 & 130 \\
         fixed direction&108 & 0.02 & 324 & 2 & 13 $\times (r+1)/3$ \\
         \hline
    \end{tabular}
    \caption{Comparing the GFs method with all variables scheme and fixed direction scheme. \#MIs represents the number of MIs of GFs.  $t_{\text{ibp}}$ represents CPU time to solve the trimmed numeric IBP system. Points represents number of numeric sample points to fit $\vec{s}$, $\epsilon$ and $\vec{x}$. Primes represents number of large prime for rational reconstruction. $t_{\text{total}}$ represents total CPU time of IBP reduction for constructing DEs of GFs. }
    \label{tab:scheme}
\end{table}

However, to reduce top-sector FIs with rank $r$, which amounts to $r+1$ FIs, in the fixed direction scheme we need to sample $\vec{c}$ for $(r+1)/3$ times. This is because we find that, for each given value of $\vec{c}$, DEs of GFs can provide 3 linearly independent relations among top-sector FIs with highest rank.
Therefore, when rank $r<29$, which is much larger than rank of FIs in usual physical problems, the efficiency of the fixed direction scheme is better.

\subsection{Comparison with constructing recurrence relations directly}

We note that using \texttt{Blade} one can reduce FIs with arbitrary powers of ISPs to lower powers, and thus construct recurrence relations directly. It is necessary to compare the efficiency between this direct way and the way based on GFs. To this end, we introduce FIs with arbitrary powers of ISPs:
\begin{align}\label{eq:n1n2}
I_{\vec{\nu}}(n_1, n_2)=\int\frac{\md^{D}\ell_1 \md^{D}\ell_2}{(\mi\pi^{D/2})^2}
\frac{\mathcal{D}_{8}^{n_1}\mathcal{D}_{9}^{n_2}}{\prod_{i=1}^7 {\cal D}_{i}^{\nu_i}}, 
\end{align}
where the ISPs $\{\mathcal{D}_{8},\ \mathcal{D}_{9}\}$ could be linear form: $\{\ell_1 \cdot k_3,\ \ell_2 \cdot k_1\}$, or quadratic form: $\{(\ell_1+k_3)^2,\ (\ell_2+k_1)^2\}$. 

Using \texttt{Blade}, we can achieve the reduction of FIs, like reducing $I_{\vec{\nu}}(n_1+1, n_2)$ and $I_{\vec{\nu}}(n_1, n_2+1)$ to $I_{\vec{\nu}}(n_1, n_2)$, which are the desired recurrence relations. The CPU time to achieve recurrence relations in this way \footnote{Although we can follow the fixed direction scheme of GF to introduce  $\widetilde{\mathcal{D}}_{9}^n=\left(c_1 {\mathcal D}_8+c_2 {\mathcal D}_9\right)^n$ in the numerator in Eq.~\eqref{eq:n1n2}, it cannot improve the efficiency because there are infinity number of singularities in the $n$ plane.} is shown in Tab.\ref{tab:ad}, which is clear less efficient comparing with the GFs method shown in Tab. \ref{tab:scheme}.
This comparison indicates the superiority of GF for generating recurrence relations.

\begin{table}[htbp]
    \begin{tabular}{|c|c|c|c|c|c|c|}
    	\hline
    	Form & \#MIs & $t_{\text{ibp}}$(s) & Points & Primes & $t_{\text{total}}$(s)\\
    	\hline
    	linear & 60 & 0.07 & 20139 & 3 & 4220\\
    	quadratic & 63 & 0.14 & 8031 & 2 & 2300 \\
    	\hline
    \end{tabular}
    \caption{Comparison of constructing recurrence relations using IBP reduction directly with different choices of ISPs (linear or quadratic form). \#MIs represents the number of MIs of FIs with general powers.  Points represents  the number of numeric sample points to fit variables $\vec{s}$, $\epsilon$ and $\vec{n}$. Other information are the same as that in Tab.\ref{tab:scheme}.}
    \label{tab:ad}
\end{table}

\sect{Two-loop double-pentagon}\label{sec:double-pentagon}

\begin{figure}[h]
    \includegraphics[width=5cm]{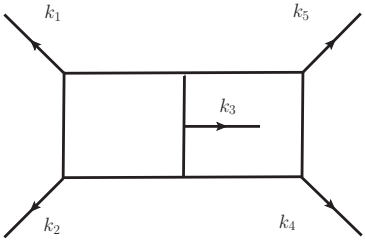}
    \caption{Two-loop double-pentagon diagram.}
    \label{fig:doublepentagon}
\end{figure}

In this section, we use the GFs method to deal with a cutting-edge problem, the massless two-loop double-pentagon family shown in Fig.\ref{fig:doublepentagon}. 
The reduction of FIs in this family up to rank 5, which are relevant for five-light-parton scattering amplitudes in QCD, has been previously accomplished
\cite{Guan:2019bcx,Klappert:2020nbg,Agarwal:2021vdh}. 
There are five external momenta $k_1$, $k_2$, $k_3$, $k_4$, $k_5$ satisfying on-shell conditions $k_i^2=0$ and momentum conservation $\sum_{i=1}^5 k_i =0$, leaving five kinematic variables $s_{12}= k_1 \cdot k_2$, $s_{23}= k_2 \cdot k_3$, $s_{13}= k_1 \cdot k_3$, $s_{14}= k_1 \cdot k_4$, $s_{34}= k_3 \cdot k_4$ as the mass scales. We choose a complete set of Lorentz scalars as
\begin{align}
\begin{split}
    &\mathcal{D}_1={\ell_1}^2,\ \mathcal{D}_2=(\ell_1+k_1)^2,\ \mathcal{D}_3={(\ell_1+k_1+k_2)}^2,\\
    &\mathcal{D}_4= {\ell_2}^2,\ \mathcal{D}_5= {(\ell_2-\ell_1)}^2,\ \mathcal{D}_6={(l_2-l_1+k_5)}^2,\\
    &\mathcal{D}_7={(\ell_2-k_3-k_4)}^2,\ \mathcal{D}_8={(\ell_2-k_4)}^2, \\
    &\mathcal{D}_9 = \ell_1 \cdot k_3,\ \mathcal{D}_{10} = \ell_2 \cdot k_1,\ \mathcal{D}_{11} = \ell_2 \cdot k_2, 
\end{split}
\end{align}
where the last three are ISPs.

\begin{table}[htbp]
    \begin{tabular}{|c|c|c|c|c|}
    \hline
          MIs & $t_{\text{ibp}}$(s) & $\text{Mem}_{\text{ibp}}$(GB) & 
          Points &
          Number of $\vec{c}$\\
         \hline
         (4,0) & 4.9 & 12.4 & 32 & $ (2+r)(1+r)/24$\\
         (0,2) & 3.6 & 2.7 & 108 & $(2+r)(1+r)/2$\\
         \hline
    \end{tabular}
    \caption{Information for reducing double-pentagon FIs 
 using  GFs method. MIs represents the max (rank,dots) of different choices of master integrals. $t_{\text{ibp}}$ represents the time spent to solve trimmed IBP system numerically for one time. $\text{Mem}_{\text{ibp}}$ represents the memory used to load and trim the raw IBP system, usually the maximal memory consumption in the calculation. Points means number of numeric sample points of $x$ to reconstruct $x$-dependence of DEs per $\vec{c}$. Number of $\vec{c}$ represents the number of samples  of $\vec{c}$ needed to reduce rank $r$ FIs.}
    \label{tab:consumption}
\end{table}

To reduce powers of ISPs, we define the following GFs
\begin{align}
G_{\vec{\nu}}^{\vec{c}}(x)=\int\frac{\md^{D}\ell_1 \md^{D}\ell_2}{(\mi\pi^{D/2})^2}
\frac{e^{x Z^{\vec{c}}}}{\prod_{i=1}^{11} {\cal D}_{i}^{\nu_i}}.
\end{align}
There are 908 MIs of these GFs in total and 18 MIs in the top sector. MIs in the top sector can be chosen as
\begin{align*}
\begin{split}
    \{&G_{(1,\dots, 1, 0, 0, 0)}^{\vec{c}},\ G_{(1,\dots, 1, -1, 0, 0)}^{\vec{c}},\ G_{(1,\dots, 1, 0, -1, 0)}^{\vec{c}},\\
    &G_{(1,\dots, 1, 0, 0, -1)}^{\vec{c}},\ G_{(1,\dots, 1, -1, -1, 0)}^{\vec{c}},\ 
    G_{(1,\dots, 1, -1, 0, -1)}^{\vec{c}},\\
    &G_{(1,\dots, 1, 0, -1, -1)}^{\vec{c}},\ G_{(1,\dots, 1, -2, 0, 0)}^{\vec{c}},\ 
    G_{(1,\dots, 1, 0, -2, 0)}^{\vec{c}},\\
    &G_{(1,\dots, 1, 0, 0, -2)}^{\vec{c}},\ G_{(1,\dots, 1, -1, -1, -1)}^{\vec{c}},\ G_{(1,\dots, 1, -2, 0, -1)}^{\vec{c}},\\
    &G_{(1,\dots, 1, -1, -2, 0)}^{\vec{c}},\ G_{(1,\dots, 1, -1, 0, -2)}^{\vec{c}},\ G_{(1,\dots, 1, 0, -2, -1)}^{\vec{c}},\\
    &G_{(1,\dots, 1, 0, -1, -2)}^{\vec{c}},\ G_{(1,\dots, 1, 0, -2, -2)}^{\vec{c}},\ G_{(1,\dots, 1, -2, -2, 0)}^{\vec{c}}\},
\end{split}
\end{align*}
which is denoted as  $(4,0)$-type because the max rank is 4 and max dot is 0. Alternatively, MIs can be chosen as
\begin{align*}
\begin{split}
    \{&G_{(1,1,1,1,1,1,1,1,0,0,0)}^{\vec{c}},\quad G_{(1,1,1,1,1,1,1,2,0,0,0)}^{\vec{c}},\\
   &G_{(1,1,1,1,1,2,1,1,0,0,0)}^{\vec{c}},\quad G_{(1,1,1,1,2,1,1,1,0,0,0)}^{\vec{c}},\\
   &G_{(1,1,1,2,1,1,1,1,0,0,0)}^{\vec{c}},\quad G_{(1,1,2,1,1,1,1,1,0,0,0)}^{\vec{c}},\\
   &G_{(1,2,1,1,1,1,1,1,0,0,0)}^{\vec{c}},\quad G_{(2,1,1,1,1,1,1,1,0,0,0)}^{\vec{c}},\\
   &G_{(1,2,1,1,1,2,1,1,0,0,0)}^{\vec{c}},\quad G_{(1,2,1,2,1,1,1,1,0,0,0)}^{\vec{c}},\\
   &G_{(1,2,2,1,1,1,1,1,0,0,0)}^{\vec{c}},\quad G_{(2,1,1,1,1,1,1,2,0,0,0)}^{\vec{c}},\\
   &G_{(2,1,1,1,1,2,1,1,0,0,0)}^{\vec{c}},\quad G_{(2,1,1,1,2,1,1,1,0,0,0)}^{\vec{c}},\\
   &G_{(2,1,1,2,1,1,1,1,0,0,0)}^{\vec{c}},\quad G_{(2,2,1,1,1,1,1,1,0,0,0)}^{\vec{c}},\\
   &G_{(1,3,1,1,1,1,1,1,0,0,0)}^{\vec{c}},\quad G_{(3,1,1,1,1,1,1,1,0,0,0)}^{\vec{c}}\},
\end{split}
\end{align*}
which is denoted as $(0,2)$-type because the max rank is 0 and max dot is 2.

Using \texttt{Blade}, we set up closed DEs of these MIs w.r.t $x$. As shown in Tab.\ref{tab:consumption}, to reduce each point (i.e., given values of kinematic variables, $\epsilon$, $\vec{c}$ and $x$), it costs $4.9~ (3.6)$ seconds and uses memory up to $12.4 ~ (2.7)$ GB, respectively for $(4,0)$-type and $(0,2)$-type. To fully construct the $x$ dependence of the DEs, we need to simulate 32 or 108 different values of $x$, respectively. Furthermore, to reduce top-sector integrals with rank $r$, we need to simulate $(2+r)(1+r)/24$ or $(2+r)(1+r)/2$ different choices of $\vec{c}$, respectively.

\begin{table}[htbp]
    \begin{tabular}{|c|c|c|c|c|}
        \hline
         rank &$t_{\text{ibp}}$(s) & $\text{Mem}_{ibp}$(GB)\\
         \hline
         4& 1.4 & 2.8\\
         5& 3.9 & 7.0\\
         6& 11 & 15.8\\
         7& 29 & 33.2\\
         8& 66 & 65.6\\
         \hline
    \end{tabular}
    \caption{Information for reducing double-pentagon FIs 
 using plain IBP method. $t_{\text{ibp}}$ represents the time spent to solve IBP system numerically using one CPU. Mem$_{ibp}$ represents the maximal memory consumption in solving the IBP system.}
    \label{tab:numIBP}
\end{table}

As a comparison, in Tab.\ref{tab:numIBP} we present the time and resource consumption for reducing FIs in the double-pentagon family with different ranks using IBP method directly. It is clear that the memory used in IBP method increases fast as the rank becoming larger. In contrast, in the GFs method, the memory is a constant value, independent of the rank of the  target FIs to be reduced. In this example, when $r>4$, the GFs method with the MIs type (0,2) needs smaller memory.

For each point, the time consumption in the GFs method is compatible with the IBP method for $r=5$. This is understandable because, to construct the DEs of GFs with the MIs type (4,0), one needs to reduce rank 5 GFs to lower ranks. However, because currently 32 points are needed to reconstruct the $x$-dependent DEs, the total CPU time $32\times 4.9\times(2+r)(1+r)/24$, is still too long. As there are only two singularities $x=0$ or $\infty$ in the $x$ plane, which are respectively regular singularity and essential singularity, it is possible to find a better set of MIs so that the DEs are in a canonical form
\begin{align} \label{eq:canonical}
    \frac{\partial }{\partial x}\vec{\cal G}^{\vec{c}}=& \left(\frac{C_{1}^{\vec{c}} }{x}+C_0^{\vec{c}}\right)\vec{\cal G}^{\vec{c}},
\end{align}
where $C_{1}^{\vec{c}}$ and $C_{0}^{\vec{c}}$ are independent of $x$. When the DEs of the MIs satisfy the canonical form, only two points of $x$ are needed to fully reconstruct the DEs instead of 32 points. In this case, the CPU time cost by the GFs method is $2\times 4.9\times(2+r)(1+r)/24$, comparable with that of IBP method when $r=7$ and the GFs method with the MIs type (4,0) becomes more efficient when $r>7$. 

Note that the DEs of sunrise family in Eq. \eqref{eq:sunriseDEs} are already in the canonical form if we set $x_i = x c_i$. For general cases, we will study the construction of canonical DEs of GFs in future publications.

\sect{Summary and outlook}\label{sec:summary}

In this paper, we have introduced a general method for constructing recurrence relations of Feynman integrals (FIs) using generating functions (GFs). We have demonstrated that by formulating and solving the differential equations (DEs) associated with GFs, we can effectively reduce FIs within a given family.

Compared to the plain IBP method, the GFs method proves to be significantly more efficient for constructing recurrence relations. When it comes to reducing FIs with specific ranks, the GFs method offers advantages in terms of time and resource consumption, particularly when the rank is large. However, for smaller ranks, the GFs method may not provide the same level of efficiency. This aspect makes the GFs method particularly appealing for addressing high-rank problems, such as  effective field theory problems, like the calculation of twist-two operator matrix elements in QCD, and  gravitational scattering problems. For example, in the calculation of Mellin Moments of splitting function by twist-two operator matrix elements, the rank of Feynman Integrals is $O(20)$ at four-loop level in planar limit \cite{Moch:2017uml} and that is $O(500)$ or higher for the full three-loop contribution \cite{Blumlein:2021enk, Blumlein:2021ryt, Gehrmann:2023ksf}. 

An important aspect to study is the selection of appropriate bases that lead to DEs of GFs in a canonical form defined in Eq.~\eqref{eq:canonical}. Techniques developed in recent years for constructing $\epsilon$-forms of FIs \cite{Henn:2013pwa} could potentially be useful in this context, although further investigation is necessary.

Given that we employ IBP relations to construct DEs of GFs, any advancements in IBP reduction techniques would also benefit our method. One possible direction is to integrate the syzygy method with the GFs method, utilizing the latter to handle the reduction of sub-sector integrals. By doing so, the DEs of GFs in the top sector can be derived with the assistance of DEs of GFs in the sub-sectors. This integration of techniques enables us to fully harness the capabilities of the GFs method, resulting in more efficient computations.

\sect{Acknowledgements}
We thank B. Feng, L.H. Huang, M. Zeng, T.Z.Yang for many useful communications and discussions.
The work was supported in part by the National Natural Science Foundation of
China (Grants No. 11875071, No. 11975029), the National Key Research and Development Program of China under
Contracts No. 2020YFA0406400, and the High-performance Computing Platform of Peking University. 
{\tt JaxoDraw}~\cite{Binosi:2003yf} was used to generate Feynman diagrams.

\providecommand{\href}[2]{#2}\begingroup\raggedright\endgroup

\end{document}